\documentclass[a4paper,12pt]{article}
\usepackage[utf8]{inputenc}
\usepackage{amsmath, amssymb, comment, hyperref, cite, graphicx, authblk}

\usepackage{anysize}
\marginsize{2.0cm}{2.0cm}{2.0cm}{2.0cm}
\begin{document}
\title{New Green's functions for some nonlinear oscillating systems and related PDEs}

\author[1]{Asatur Zh. Khurshudyan\footnote{Email: khurshudyan@mechins.sci.am}}

\affil[1]{Department on Dynamics of Deformable Systems and Coupled Fields, Institute of Mechanics, NAS of Armenia}

\date{}

\maketitle

\begin{abstract}
During the past three decades, the advantageous concept of the Green's function has been extended from linear systems to nonlinear ones. At that, there exist a rigorous and an approximate extensions. The rigorous extension introduces the so-called backward and forward propagators, which play the same role for nonlinear systems as the Green's function plays for linear systems. The approximate extension involves the Green's formula for linear systems with a Green's function satisfying the corresponding nonlinear equation. For the numerical evaluation of nonlinear ordinary differential equations the second approach seems to be more convenient. In this article we study a hierarchy of nonlinear \emph{partial} differential equations that can be approximated by the second approach. Green's functions for particular non-linearities are derived explicitly. Numerical error analysis in the case of exponential non-linearity for different source functions supports the advantage of the approach.


\end{abstract}

\section{Introduction}

Since its invention, the Green's function technique was used to analyze numerous \emph{linear} systems in various areas of natural science. The key pillar of the Green's function technique is the superposition principle, which makes the technique to be applicable exceptionally for linear systems. Nevertheless, two extensions of the Green's function technique to the \emph{nonlinear} systems has been reported in the past three decades. More specifically, in \cite{Cacuci1988, Cacuci1989, Cacuci1990} (see also other articles by D. Cacuci) the concept of the so-called backward and forward propagators is introduced for general nonlinear systems. It was established that the propagators play the same role for nonlinear systems as the Green's function does for linear systems. The approach has been tested for main types of nonlinear systems. The second approach is due to M. Frasca \cite{Frasca2007, Frasca2008}, who has shown the advantages of the approximate extension in quantum field theory. The solution of second order nonlinear ordinary differential equations of a specific form (we formally call it ``oscillating'' equation, since the leading order term is the second order derivative of the state function with respect to time) is approximated by a formula similar to the Green's representation formula for linear equations with the difference that the Green's function is determined from the corresponding \emph{nonlinear} equation. It is numerically established that the proposed formula provides a low-error approximation for two particular non-linearities.

Such extensions play into the hands of physicists, since they considerably simplify the analysis of the dependence of the unknown function on internal and external parameters and the establishment of key features that are somehow hidden in intermediate steps of numerical computations in the case of nonlinear processes \cite{Flesch1987, Schwartz1997, Qiao2004, Guo2016, Gao2017}.

Several very specific non-linearities supporting the Frasca's approximation formula are considered in this paper. All the systems are shown to be the reduced versions of nonlinear partial differential equations arising in different areas of natural science, ranging from vibrarions in solids and fluids, quantum field theory to biology. Numerical analysis establishes the advantage of the approximation formula and opens new perspectives of its practical use. Note also that the idea of \cite{Avetisyan2017} can be applied on the results of this paper, in order to consider control problems for new ``oscillating'' nonlinear systems.

\section{Frasca's representation formula}

In \cite{Frasca2007, Frasca2008} it is shown that for a generic function $N$ and a given source function $f$, the nonlinear ordinary differential equation
\begin{equation}\label{gennonlin}
\frac{d^2 w}{d t^2} + N\left(w, t\right) = f\left(t\right),~~ t > 0,
\end{equation}
admits the approximate solution
\begin{equation}\label{Frascaapprox}
w\left(t\right) \approx \int_0^t G\left(t, \tau\right)f\left(\tau\right) {\rm d}\tau.
\end{equation}
Here $G$ is the formal Green's function of (\ref{gennonlin}), i.e. the solution of
\begin{equation}\label{Greeneq}
\frac{d^2 G}{d t^2} + N\left(G, t\right) = s_1 \delta\left(t - \tau\right),~~ t > 0,~ \tau > 0,
\end{equation}
in the sense of distributions, where $\delta$ is the Dirac distribution. Homogeneous Cauchy conditions must be attached to (\ref{Greeneq}). Here $s_1$ is a real parameter that must be chosen to minimize the approximation error.

Two particular forms of $N$ are considered in \cite{Frasca2007} providing an exact solution of (\ref{Greeneq}). Specifically, if
\[
N\left(w, t\right) = w^3,
\]
then the nonlinear Green's function has the form
\[
G\left(t, \tau\right) = 2^{\frac{1}{4}}\theta\left(t - \tau\right) \cdot \operatorname{sn}\left[\frac{t - \tau}{2^{\frac{1}{4}}},i\right].
\]
Here $\theta$ is the Heaviside function, and $\operatorname{sn}$ is the Jacobi snoidal function.

Furthermore,
\[
N\left(w, t\right) = \sin w
\]
admits the exact solution
\[
G\left(t, \tau\right) = 2\theta\left(t - \tau\right) \cdot \operatorname{am}\left[\frac{t - \tau}{\sqrt{2}}, \sqrt{2}\right].
\]
Here $\operatorname{am}$ is the Jacobi amplitude function. Both cases provide low-error application of (\ref{Frascaapprox}) for particular source functions $f$.

\subsection{PDE reduction via generalized variable separation}

The system (\ref{gennonlin}) is a reduced version of nonlinear partial differential equations that are linear in
\[
\frac{\partial^2 \tilde{w}}{\partial t^2}.
\]
Indeed, using the generalized separation of variables (see \cite{Polyanin2000, Barannyk2011} for details), the equation
\begin{equation}\label{nonlinwave}
\frac{\partial^2 \tilde{w}}{\partial t^2} + \tilde{N}\left(\frac{\partial^k \tilde{w}}{\partial x^k}, x, t\right) = \tilde{f}\left(x, t\right), ~~ k = 0, 1, \dots,
\end{equation}
can be reduced to (\ref{gennonlin}). Consider, for instance, the one-dimensional nonlinear wave equation
\[
\frac{\partial^2 \tilde{w}}{\partial t^2} = \alpha \frac{\partial}{\partial x}\left[\exp\left[\lambda x\right] \frac{\partial \tilde{w}}{\partial x}\right] + \tilde{N}\left(w, x, t\right),
\]
with a general non-linearity $N$ and real parameters $\alpha$ and $\lambda$. Using the method of generalized variable separation and introducing the auxiliary variable
\begin{equation}\label{gensep}
\chi^2 = a_1\left[ \frac{\exp\left[- \lambda x\right]}{\alpha \lambda^2} - \frac{\left(t + a_2\right)^2}{4} \right],
\end{equation}
where $a_1$ and $a_2$ are arbitrary constants, the above wave equation is reduced to the nonlinear ordinary differential equation
\[
\frac{d^2 w}{d \chi^2} + \frac{4}{a_1} N\left(w, \chi\right) = 0.
\]

Note that equation (\ref{nonlinwave}) arises in many areas of physics including gravity, quantum field theory, engineering and fluid mechanics describing, as a rule, nonlinear wave phenomena in solids or fluids \cite{Drumheller1998}, as well as in biology \cite{Murray2002}.

\section{New cases of validity of (\ref{Frascaapprox})}

Using the Frasca's idea, nonlinear Green's functions are derived in this section for particular non-linearities $N$. In what follows, $c_1$ and $c_2$ are integration constants determined from homogeneous Cauchy conditions

Assume that
\[
N\left(w, t\right) = w^2.
\]
This case corresponds to the quadratic Gordon equation transformed by (\ref{gensep}). Then (\ref{Greeneq}) accepts the exact solution
\[
G\left(t, \tau\right) = -\frac{1}{c} \theta\left(t - \tau\right) \wp\left(c \left(t - \tau\right) + c_1; 0, c_2\right), ~~ c = \left(-6\right)^{-\frac{1}{3}}.
\]
Here $\wp$ is the Weierstrass elliptic function
\[
\wp\left(t; \omega_1, \omega_2\right) = \frac{1}{t^2} + \sum_{n^2 + m^2 \neq 0}\left[\frac{1}{\left(t + \omega_1 m + \omega_2 n\right)^2} - \frac{1}{\left(\omega_1 m + \omega_2 n\right)^2}\right].
\]

It is possible to derive an exact solution also in the case when
\[
N\left(w, t\right) = \frac{1}{w}.
\]
In this case, (\ref{Greeneq}) admits the exact solution
\[
G\left(t, \tau\right) = c_1 \theta\left(t - \tau\right) \exp\left[ - \varphi^2\left(t - \tau; c_1, c_2\right) \right],
\]
where
\[
\varphi\left(t; c_1, c_2\right) = \operatorname{erf}^{-1}\left[-\sqrt{\frac{2}{\pi}} \left|c_1\right|\left|t + c_2\right|\right],
\]
$\operatorname{erf}^{-1}$ is the inverse of the Gauss error function
\[
\operatorname{erf}\left(t\right) = \frac{2}{\sqrt{\pi}}\int_0^t \exp\left[-\tau^2\right] {\rm d}\tau.
\]

In the case of the exponential nonlinearity
\begin{equation}\label{expnonlin}
N\left(w, t\right) = \exp w,
\end{equation}
Eq. (\ref{Greeneq}) admits the exact solution
\[
G\left(t, \tau\right) = \theta\left(t - \tau\right) \cdot \ln\left[\frac{1}{2}c_1\left(1 - \tanh^2\left[\frac{1}{2}\sqrt{c_1 \left(t + c_2\right)^2}\right]\right)\right].
\]

It is also possible to consider non-linearities containing differentials of the unknown function of the form
\[
N = N\left(\frac{d w}{dt}, w, t\right).
\]
In particular, if
\[
N\left(\frac{d w}{dt}, w, t\right) = w \frac{d w}{dt},
\]
the exact solution of (\ref{Greeneq}) reads as
\[
G\left(t, \tau\right) = c_1 \theta\left(t - \tau\right) \cdot \tanh\left[\frac{1}{2}c_1\left(t + c_2\right)\right].
\]

There exist numerous other non-linearities, for which the nonlinear Green's function can be found in explicit form (see the handbook \cite{Polyanin2017}).

\subsection{Wave equation with exponential non-linearity}

Let us study the case of the exponential non-linearity (\ref{expnonlin}). Consider the one-dimensional wave equation
\begin{equation}\label{waveexpnonlin}
\frac{\partial^2 \tilde{w}}{\partial t^2} = \frac{\partial}{\partial x}\left[\exp\left(\lambda x\right) \frac{\partial \tilde{w}}{\partial x}\right] + \exp \tilde{w} + \tilde{f}\left(x, t\right), ~~ -\infty < x < \infty, ~ t > 0,
\end{equation}
describing the nonlinear wave propagation in inhomogeneous media. For simplicity, set $\lambda = 2$, $a_1 = 4$ and $a_2 = 0$, so that the transformation (\ref{gensep})
\[
\chi^2 = \exp\left(-2x\right) - t^2,
\]
reduces the wave equation to the ordinary differential equation
\[
\frac{d^2 w}{d \chi^2} + \exp\left[w\left(\chi\right)\right] = f\left(\chi\right).
\]
According to the Frasca's approach, its solution can be approximated as follows:
\[
w\left(\chi\right) \approx \int G\left(\chi - \zeta\right) f\left(\zeta\right) {\rm d}\zeta.
\]
Here the Green's function is determined from (\ref{Greeneq}) under corresponding homogeneous Cauchy conditions as follows:
\[
G\left(\chi\right) = \theta\left(\chi\right) \ln\left[1 - \tanh^2\frac{\chi}{\sqrt{2}}\right].
\]
Thus, the general solution of the wave equation (\ref{waveexpnonlin}) can be approximated by
\[
w\left(x, t\right) \approx \int_{-\infty}^{\infty} \int_0^t \tilde{G}\left(x - \xi, t - \tau\right) f\left(\xi, \tau\right) {\rm d}\xi {\rm d}\tau,
\]
where
\[
\tilde{G}\left(x, t\right) = G\left(\chi\left(x, t\right)\right).
\]

\section{Numerical vs exact solutions}

We slightly modify the Frasca's approximation formula (\ref{Frascaapprox}) by introducing a new scale parameter $s_1$ according to
\begin{equation}\label{Frascaapproxs1}
w\left(t\right) \approx s_2 \int_0^t G\left(t, \tau\right)f\left(\tau\right) {\rm d}\tau.
\end{equation}
In order to check the applicability of (\ref{Frascaapproxs1}), a comparison between the derived exact and numerical solutions is provided for particular non-linearities $N$ and source functions $f$. Let the non-linearity is given by (\ref{expnonlin}). To measure the error between the approximate solution $w_{\rm app}$ and the exact solution $w_{\rm exact}$, the logarithmic error function
\[
\operatorname{Er}\left(t\right) = \log_{10}\left|w_{\rm app}\left(t\right) - w_{\rm exact}\left(t\right)\right|
\]
is quantified, evaluating the difference between the two solutions in degrees of $10$.

\begin{figure}[ht]
\centerline{\includegraphics[width = 3in]{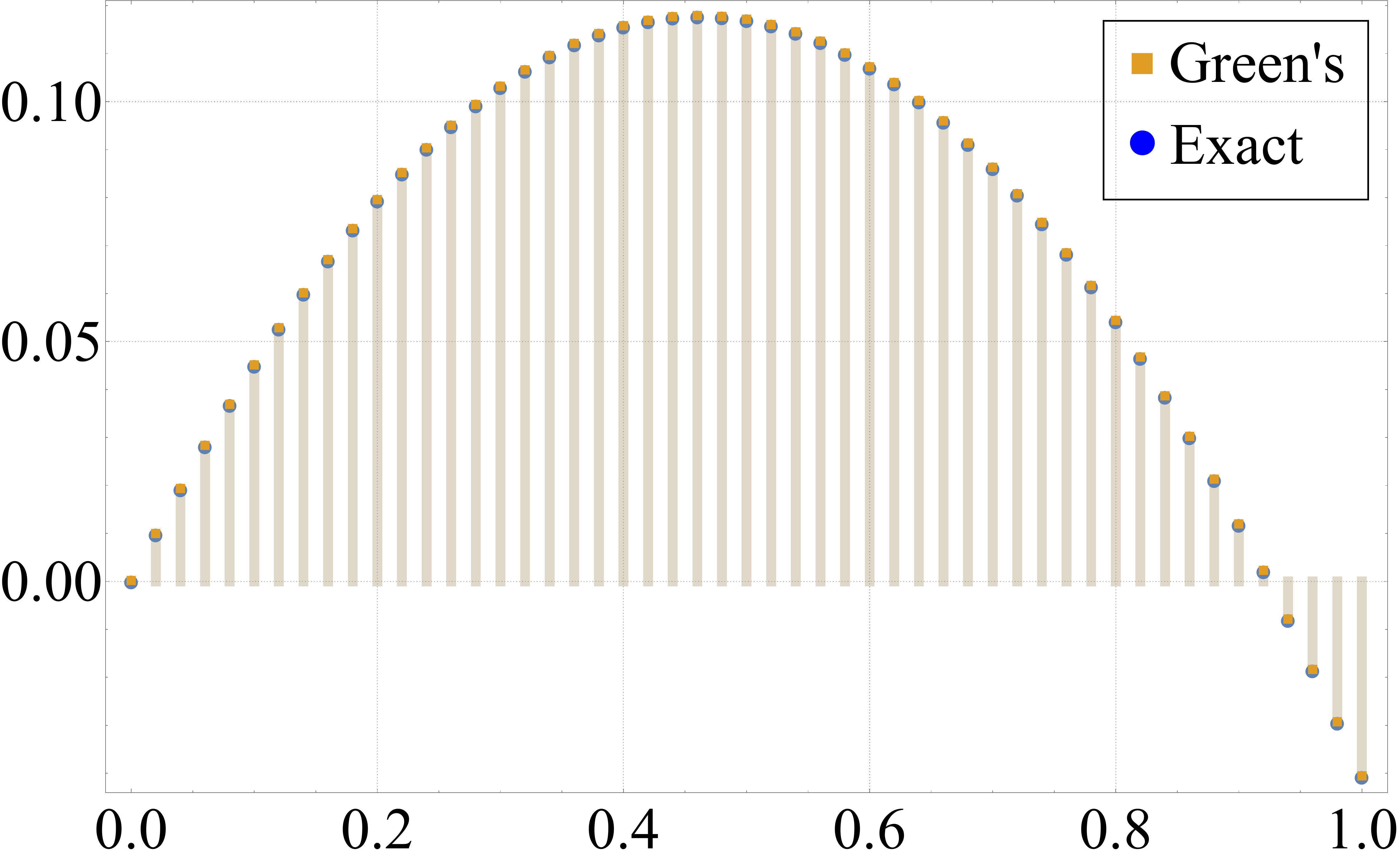} ~ \includegraphics[width=3in]{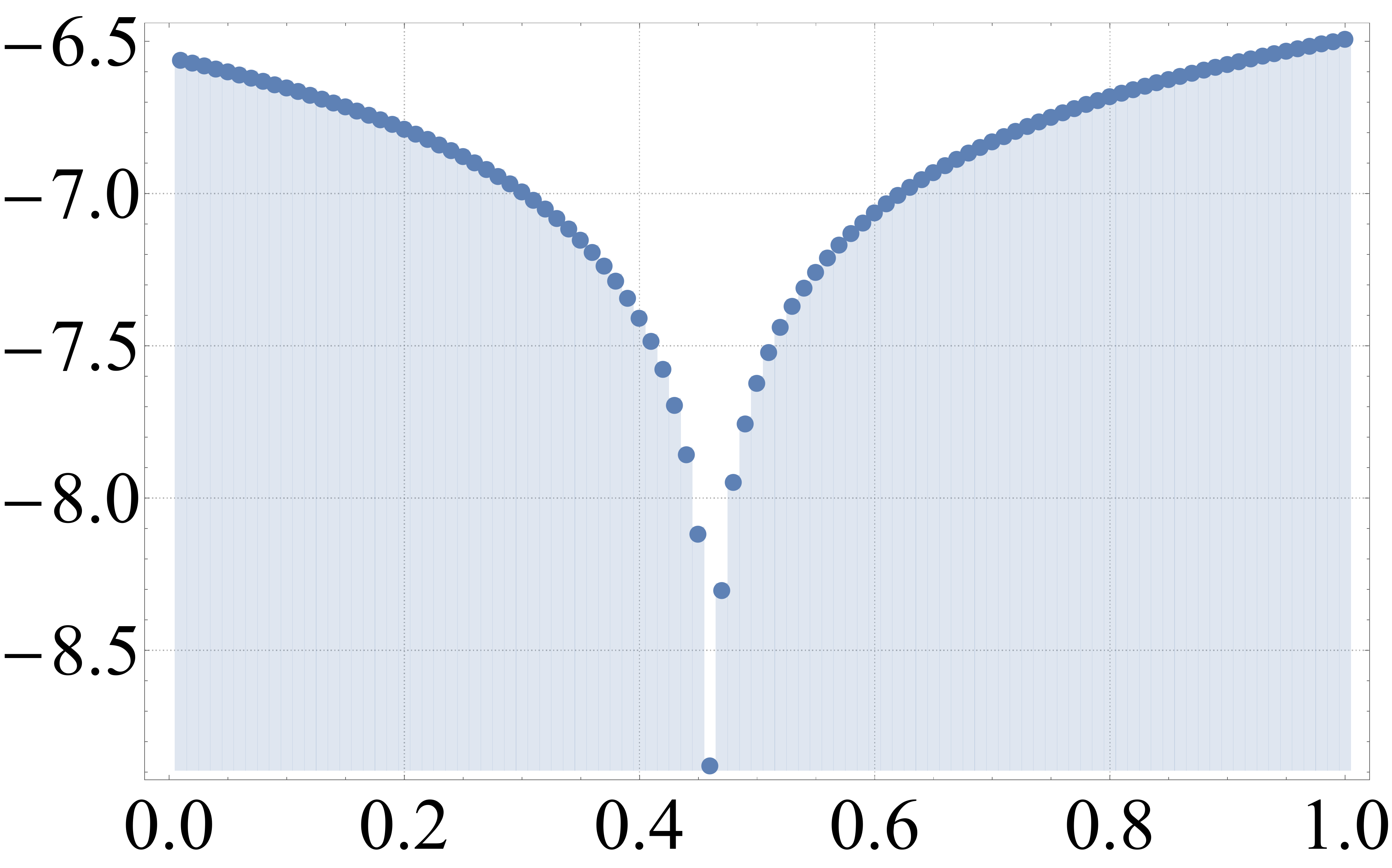}}
\vspace*{8pt}
\caption{Discrete plot of exact and approximate solutions (left) and $\operatorname{Er}$ (right) for $f\left(t\right) = \delta\left(t\right)$
\label{fig1}}
\end{figure}

\begin{figure}[ht]
\centerline{\includegraphics[width = 3in]{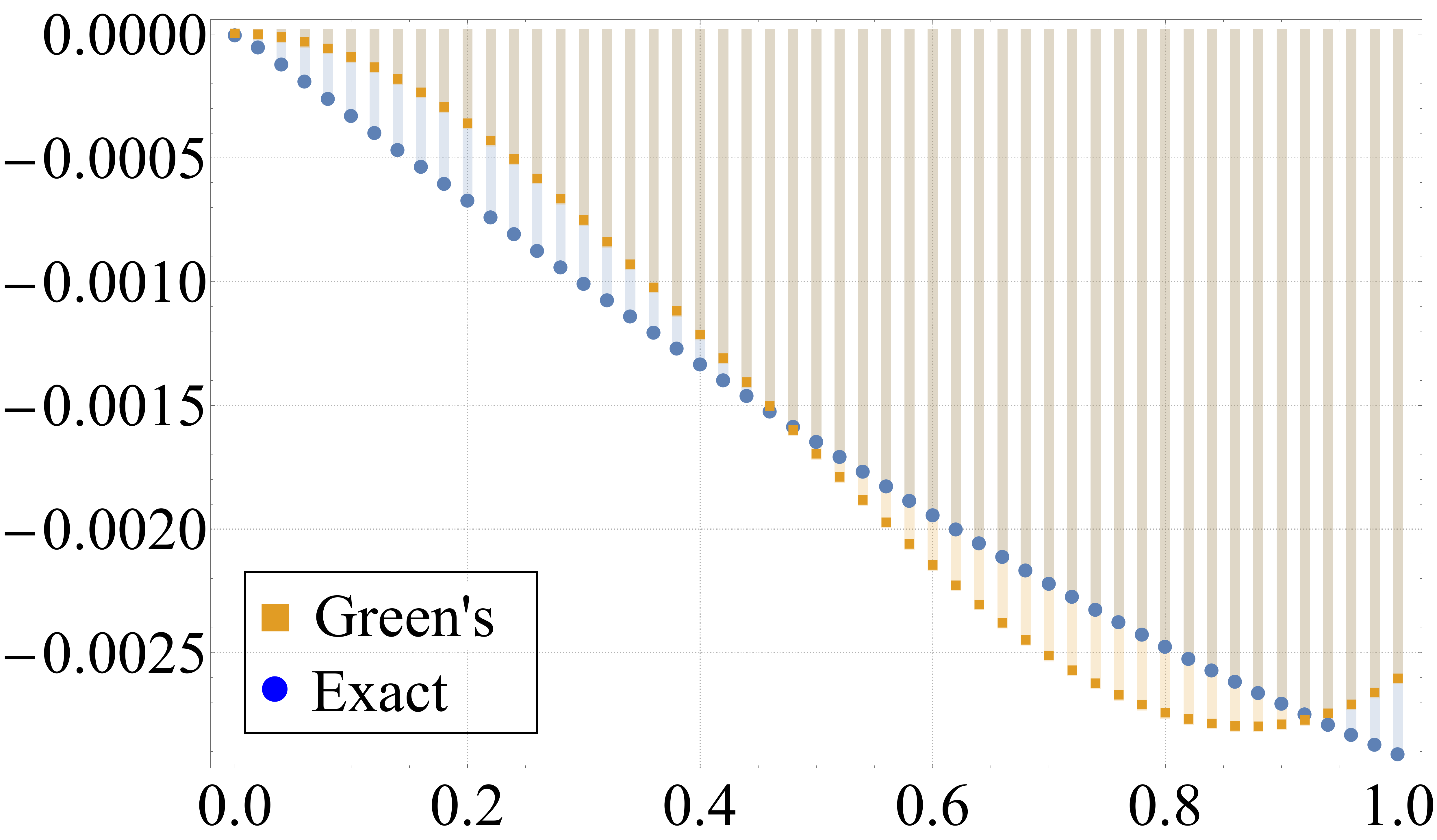} ~ \includegraphics[width=3in]{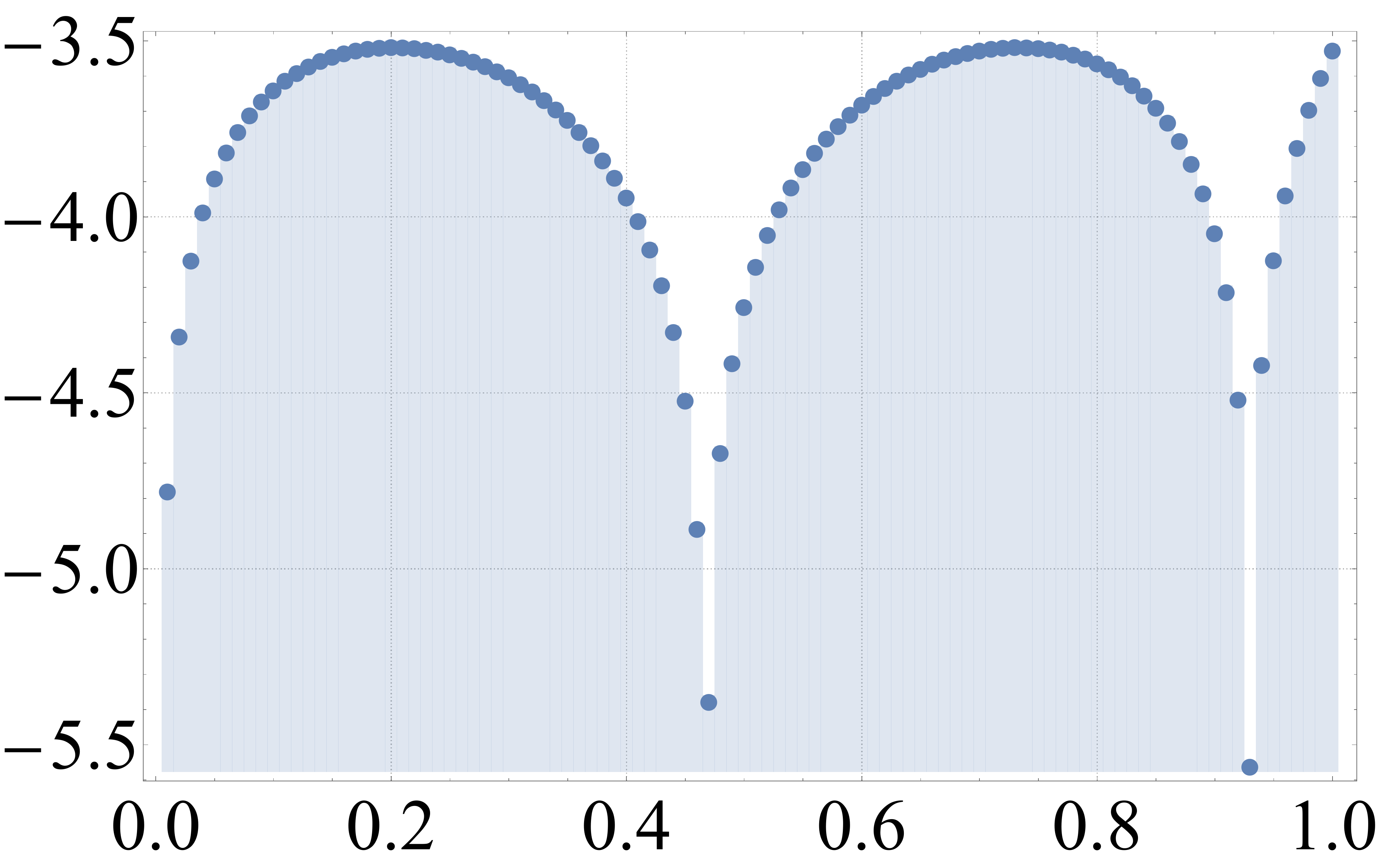}}
\vspace*{8pt}
\caption{Discrete plot of exact and approximate solutions (left) and $\operatorname{Er}$ (right) for $f\left(t\right) = \theta\left(t\right)$
\label{fig2}}
\end{figure}

\begin{figure}[ht]
\centerline{\includegraphics[width = 3in]{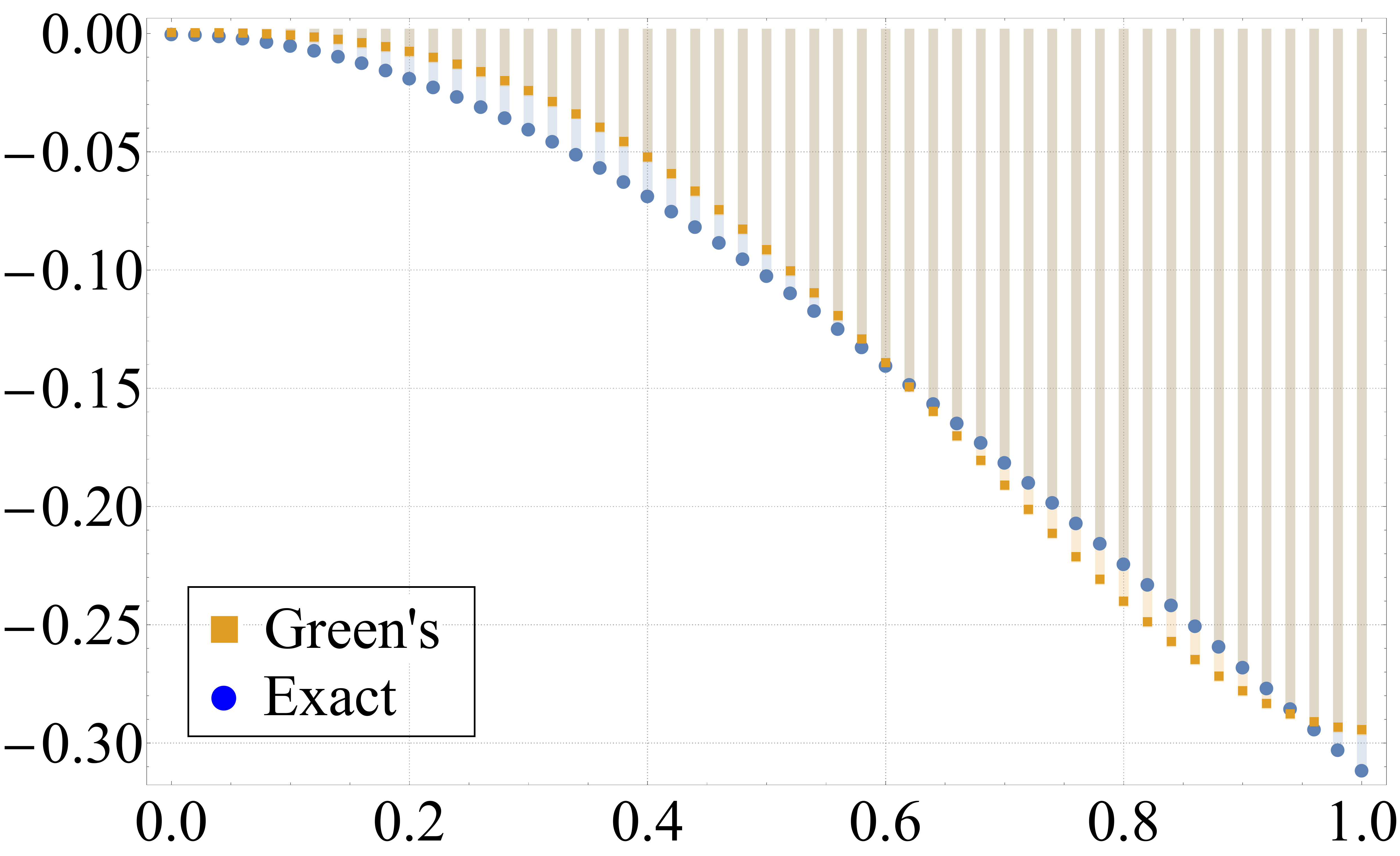} ~ \includegraphics[width=3in]{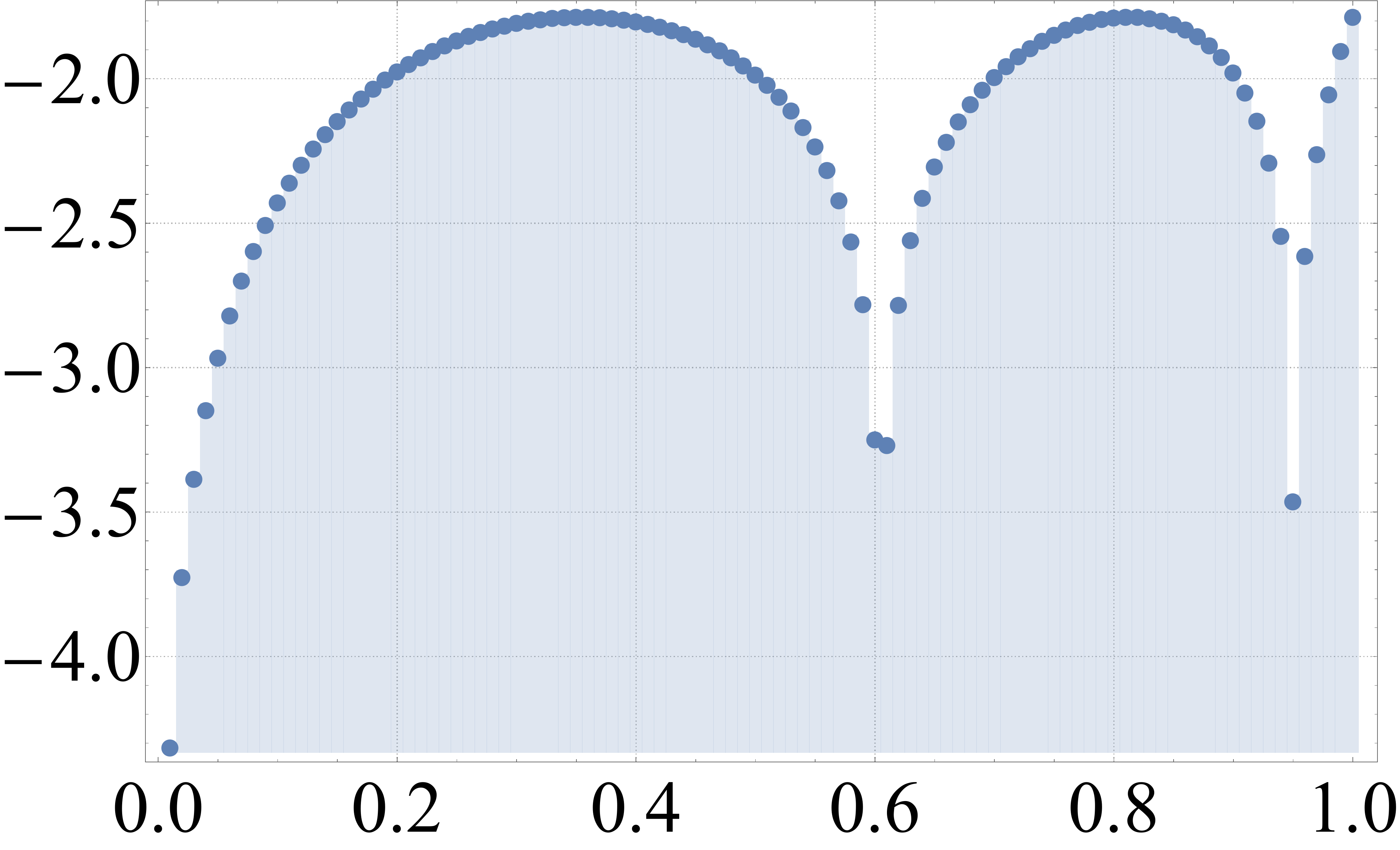}}
\vspace*{8pt}
\caption{Discrete plot of exact and approximate solutions (left) and $\operatorname{Er}$ (right) for $f\left(t\right) = \sin\left(t\right)$
\label{fig3}}
\end{figure}

\begin{figure}[ht]
\centerline{\includegraphics[width = 3in]{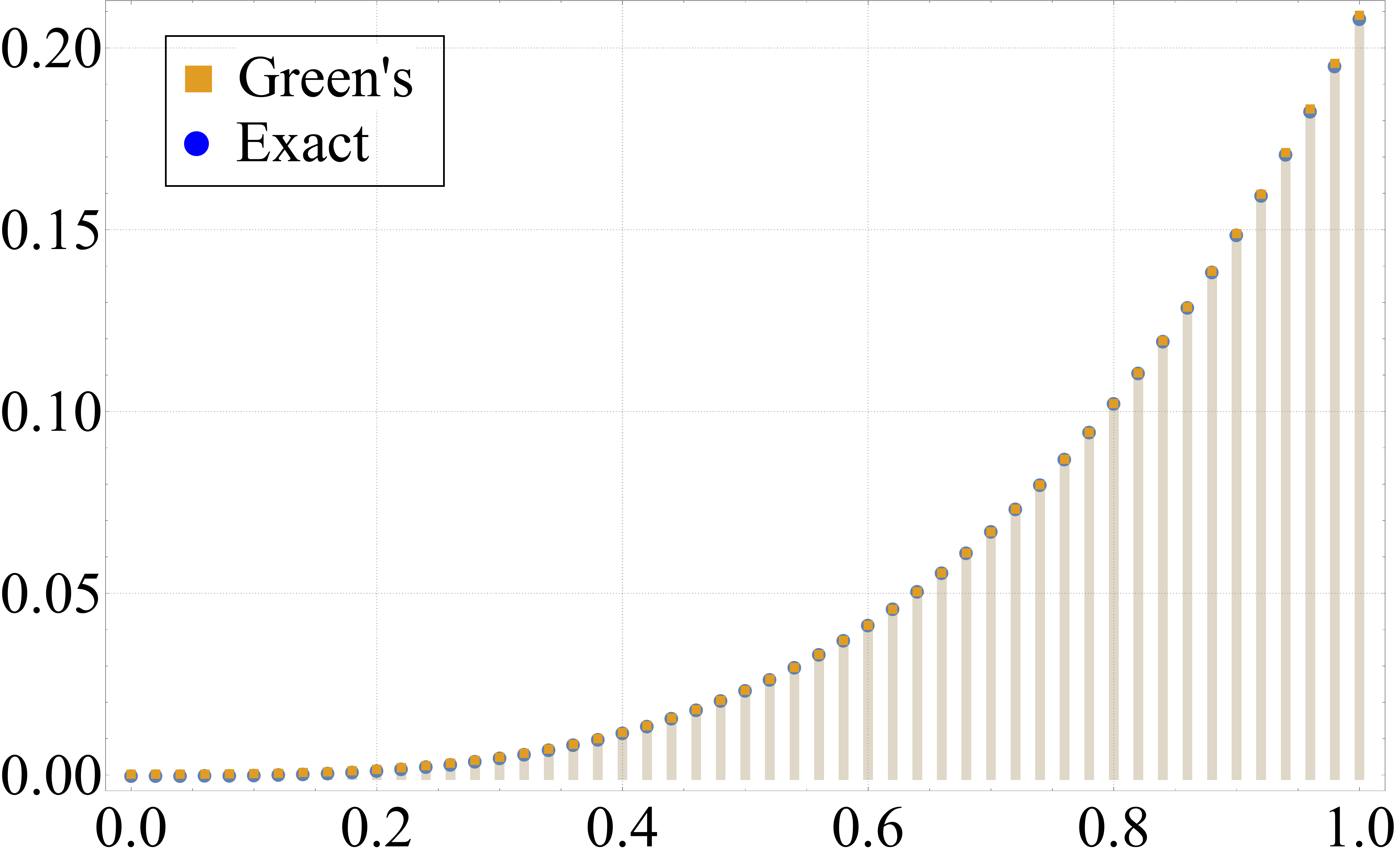} ~ \includegraphics[width=3in]{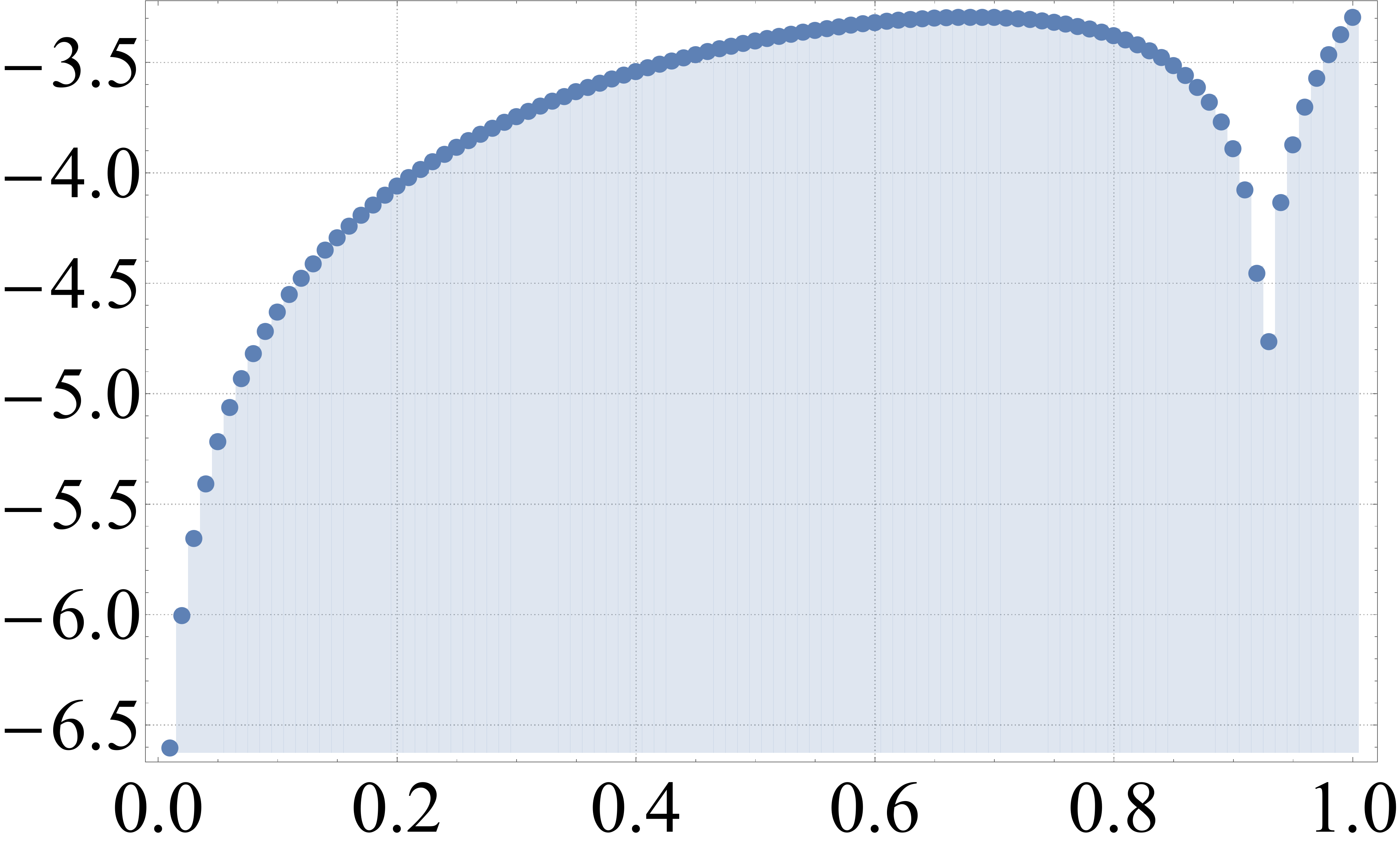}}
\vspace*{8pt}
\caption{Discrete plot of exact and approximate solutions (left) and $\operatorname{Er}$ (right) for $f\left(t\right) = \exp\left(t\right)$
\label{fig4}}
\end{figure}

\begin{figure}[ht]
\centerline{\includegraphics[width = 3in]{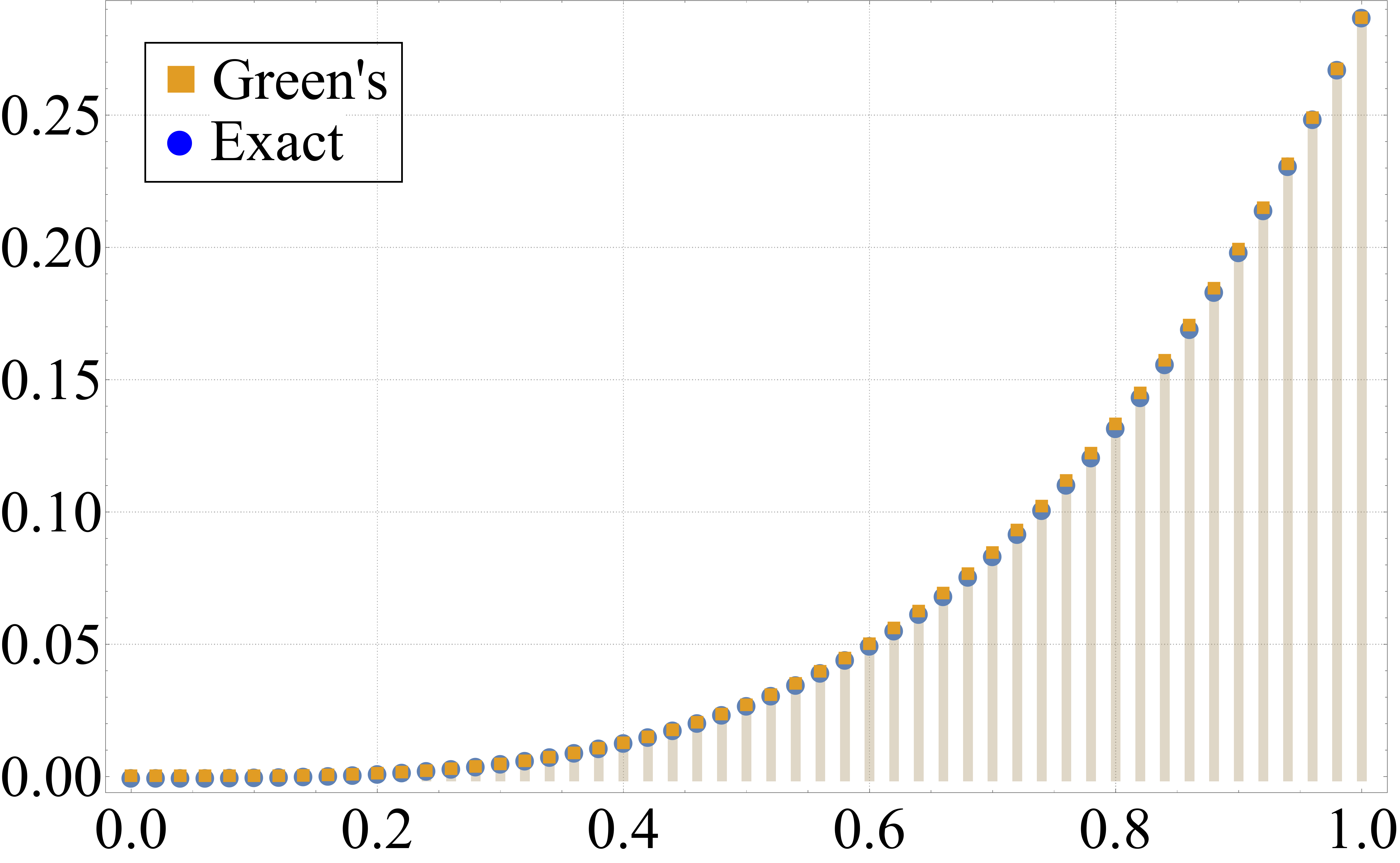} ~ \includegraphics[width=3in]{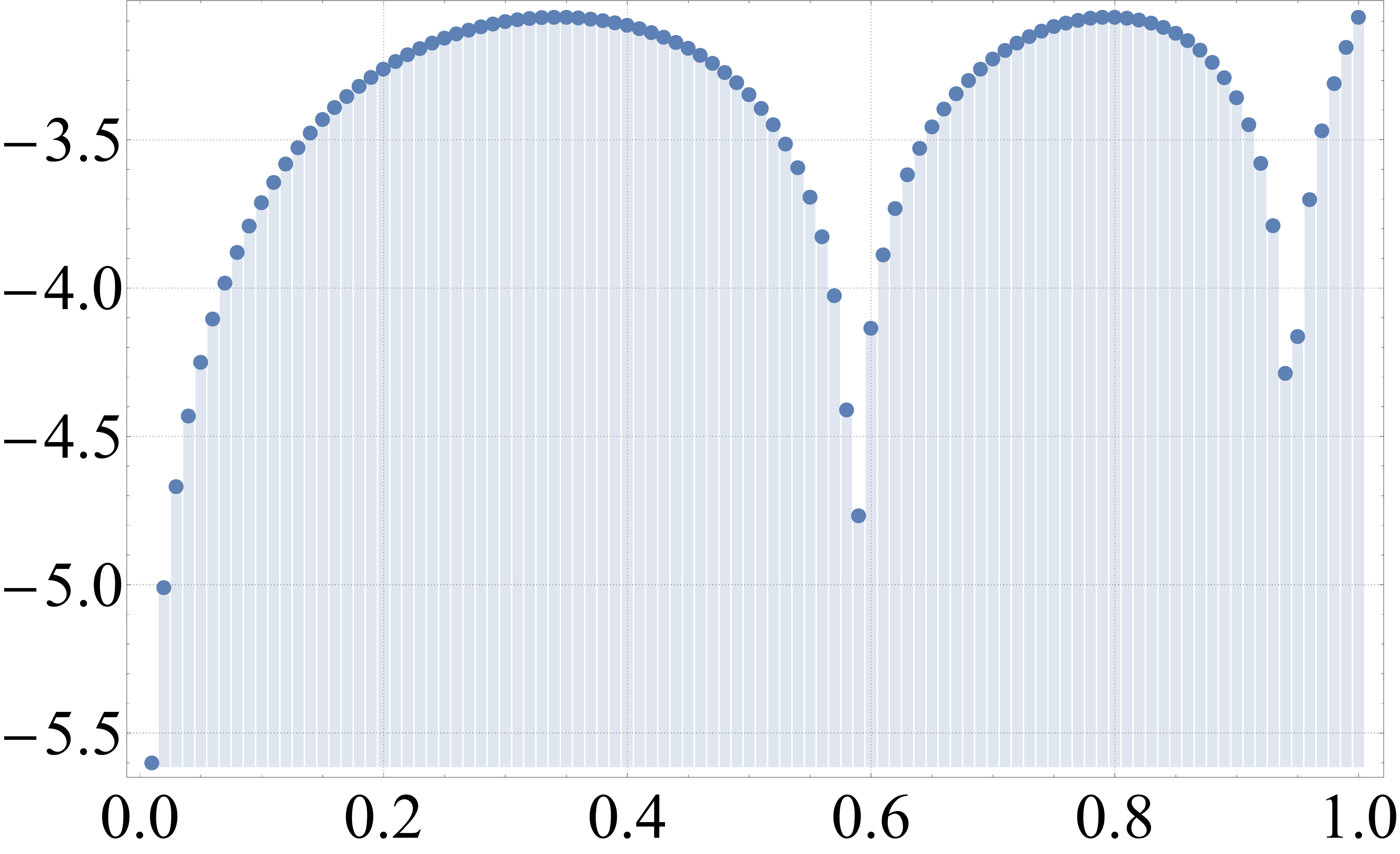}}
\vspace*{8pt}
\caption{Discrete plot of exact and approximate solutions (left) and $\operatorname{Er}$ (right) for $f\left(t\right) = 1 + t + t^2 + t^3$
\label{fig5}}
\end{figure}

\begin{figure}[ht]
\centerline{\includegraphics[width = 3in]{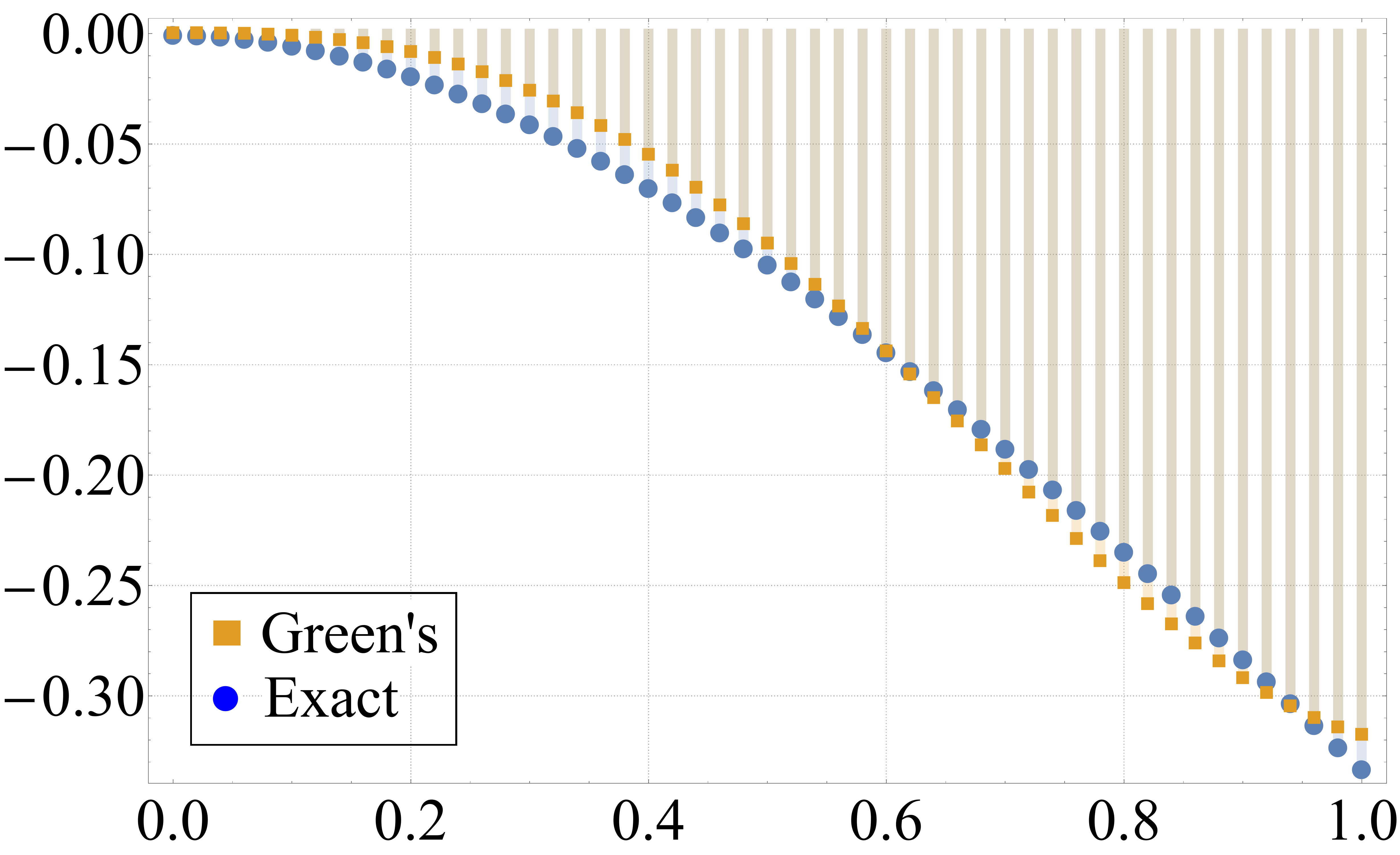} ~ \includegraphics[width=3in]{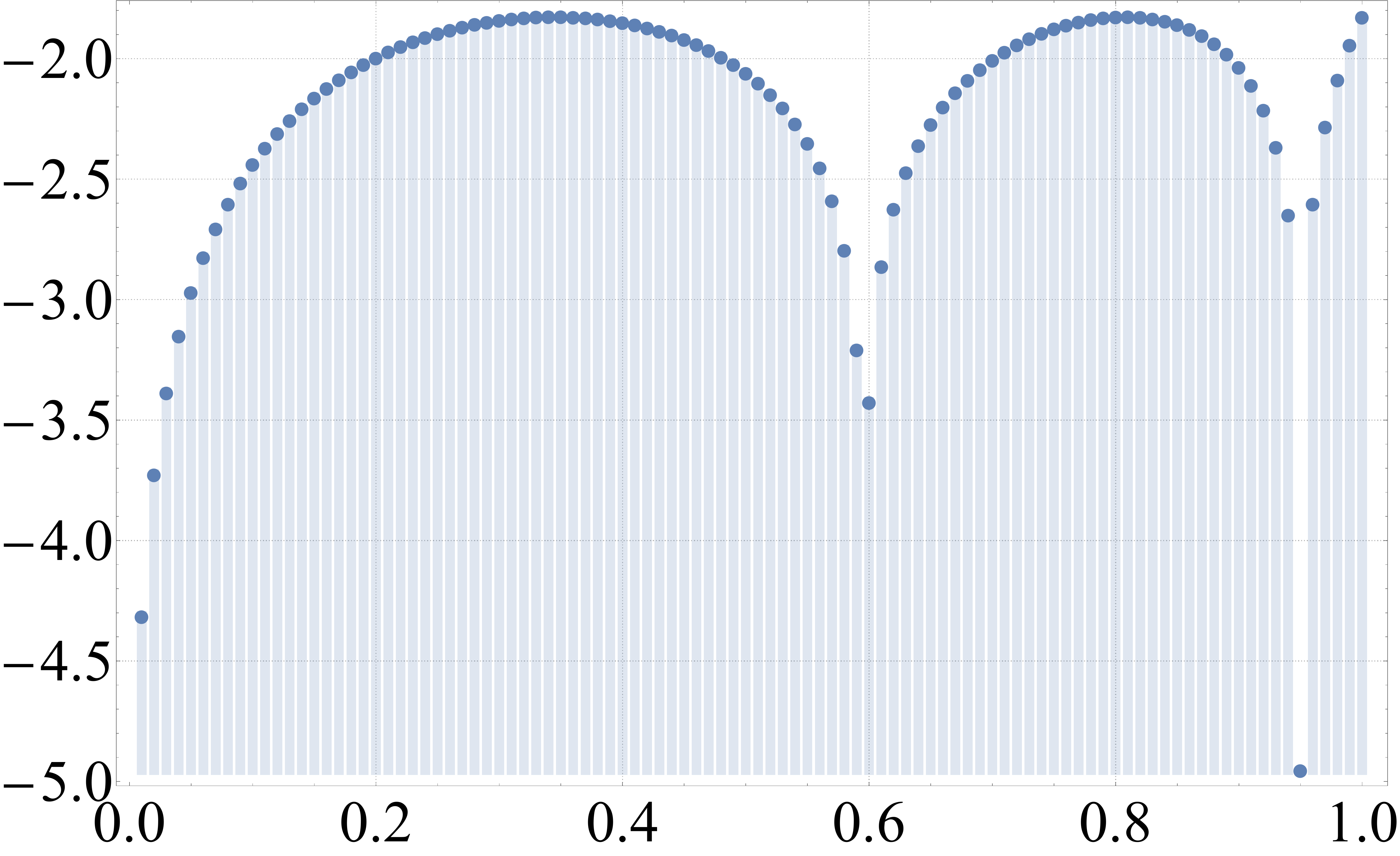}}
\vspace*{8pt}
\caption{Discrete plot of exact and approximate solutions (left) and $\operatorname{Er}$ (right) for $f\left(t\right) = \ln\left(1 + t\right)$
\label{fig6}}
\end{figure}

First, we consider the case when $f\left(t\right) = \delta\left(t\right)$. It is evident from Fig. \ref{fig1} that in this case the formula (\ref{Frascaapproxs1}) provides a very good approximation. Less accurate but still efficient approximation is observed for various other right hand sides, including switching, trigonometric, exponential, polynomial and logarithmic functions (see Figs. \ref{fig2}--\ref{fig6}).

Errors of numerical approximation of the solution of (\ref{gennonlin}) by (\ref{Frascaapproxs1}) in the case of (\ref{expnonlin}) and different source influences $f$ is brought in Table \ref{tab1}.

\begin{table}[ht]
\centering
\begin{tabular}{ c c c c c}
 $f\left(t\right)$ & $\min\operatorname{Er}$ & $\max\operatorname{Er}$ & $s_1$ & $s_2$ \\ \hline
 
 $\delta\left(t\right)$ & $-9$ & $-6.5$ & $1$ & $2$ \\ \hline
 
 $\theta\left(t\right)$ & $-6$ & $-3.5$ & $0.93107$ & $-0.047109$ \\ \hline
 
 $\sin\left(t\right)$ & $-4.5$ & $-1.75$ & $0.72126$ & $-19.2534$ \\ \hline
 
 $\exp\left(t\right)$ & $-6.5$ & $-3.25$ & $0.01$ & $-1.0142$ \\ \hline
 
 $1 + t + t^2 + t^3$ & $-5.6$ & $-3$ & & $0.07149$ $-1.45421$ \\ \hline

 $\ln\left(1 + t\right)$ & $-5$ & $-1.75$ & $-20$ & $0.7743$ \\ \hline 
\end{tabular}
\caption{Minimal and maximal logarithmic errors of approximation for various source functions: exponential non-linearity}
\label{tab1}
\end{table}

The efficiency of (\ref{Frascaapproxs1}) is observed for other types of nonlinearities as well.

\newpage\clearpage

\section*{Conclusion}

The validity of the Frasca's representation formula for ordinary differential equations is shown numerically for some new classes of nonlinear equations. At this, a certain link between nonlinear partial differential equations describing nonlinear wave phenomena and ordinary differential equations for which the Frasca's formula is true, is established through the generalized separation of variables. The numerical solution represented in terms of the nonlinear Green's function is compared with its exact solution in the case of exponential non-linearity. Consideration of different types of source functions makes the comparison more trusty.

It is observed that the error strongly depends on the source function. Fixing the scaling parameter occurring in the right-hand side of the nonlinear equation for the Green's function, the approximation error is minimized with respect to the second parameter occurring in the Frasca's approximation formula. The simultaneous minimization of the approximation error with respect to both parameters remains challenging.

{\bf Acknowledgments}. We highly appreciate the constructive remarks of the referees helped us to improve the manuscript.

\end{document}